\documentclass[12pt]{article}

\usepackage{times,epsfig,soul,graphics,amssymb,latexsym,amsmath,setspace,fullpage,enumerate,placeins,float}
\usepackage{array,threeparttable,hyperref,url}
\usepackage[usenames]{color}
\usepackage{booktabs, comment}
\usepackage{multirow}
\usepackage{algorithmicx,verbatim}
\usepackage[affil-it]{authblk}
\usepackage{graphicx}
\usepackage[colorinlistoftodos]{todonotes}
\usepackage[super,sort&compress]{natbib}
\setcitestyle{comma,numbers,super,open={},close={}} 
\usepackage{hyperref}
\usepackage{authblk}
\usepackage{courier}
\usepackage{amsmath}
\usepackage{listings}
\usepackage{minted}
\usepackage{comment}
\usepackage{subcaption}
\title{A Targeted Learning Framework for Estimating Restricted Mean Survival Time Difference using Pseudo-observations}
\author{Man Jin\thanks{Address: 1 North Waukegan Rd, North Chicago, IL 60064; {email: manmandy.jin@abbvie.com}. \ \ This article has been submitted to a jounal and is under review.}, \ \ Yixin Fang}
\affil{Data and Statistical Sciences, AbbVie Inc.}
\date{\today}
\date{}

\usepackage{natbib}
\usepackage{graphicx}
\usepackage{amssymb}
\usepackage{setspace}

\DeclareMathAlphabet\mathbfcal{OMS}{cmsy}{b}{n}

\def\boxit#1{\vbox{\hrule\hbox{\vrule\kern6pt\vbox{\kern6pt#1\kern6pt}\kern6pt\vrule}\hrule}}

\allowdisplaybreaks

\begin{document}

\maketitle

\begin{abstract}

A targeted learning (TL) framework is developed to estimate the difference in the restricted mean survival time (RMST) for a clinical trial with time-to-event outcomes. The approach starts by defining the target estimand as the RMST difference between investigational and control treatments. Next, an efficient estimation method is introduced: a targeted minimum loss estimator (TMLE) utilizing pseudo-observations. Moreover, a version of the copy reference (CR) approach is developed to perform a sensitivity analysis for right-censoring. The proposed TL framework is demonstrated using a real data application. 

\end{abstract}

{\it Keywords: Censoring; Estimand; RMST; Sensitivity analysis; Time-to-event}

\section{Introduction}

The targeted learning (TL) framework has been applied to both experimental and observational studies \cite{van2011targeted}. In this paper, we introduce a specialized version of the TL framework to estimate the effect of an investigational treatment compared to a control treatment for time-to-event outcomes.

Consider a randomized controlled trial (RCT) or a cohort study with a time-to-event outcome. Let $A$ indicate the assignment of the treatment group: $A=1$  for treatment and $A=0$ for control. Let $X$ represent a vector of baseline covariates collected at time zero, and let $T$ denote the time to the event of interest (e.g. time to death) and $T=0$ as the start of treatment.

Let $T^{a=1}$ and $T^{a=0}$ be the potential outcomes had the subject been treated with treatment $1$ and treatment $0$, respectively. Assume that the target estimand is defined in terms of the difference in the restricted mean survival time (RMST)\cite{irwin1949standard, karrison1987restricted}, 
\begin{align}
    \theta=\mathbb{E}(T^{a=1}\wedge \tau)-\mathbb{E}(T^{a=0}\wedge \tau),\label{estimand}
\end{align}
for a given cutoff time $\tau$, where the expectation is over a population.

There is an alternative way to express the target estimand $\theta$. Let $S^a(t)=\mathbb{P}(T^a>t)$ be the survival function of $T^a$ and let $\mu_a=\mathbb{E}(T^a\wedge \tau)$. Thus, 
\begin{align}
  \mu_a=\mathbb{E}(T^a\wedge \tau)=\int_0^{\tau} S^a(t)dt,  
\end{align}
which is the area under the survival function on $[0, \tau]$.
Therefore, 
\begin{align}
    \theta=\mu_1-\mu_0=\int_0^{\tau} S^{a=1}(t)dt-\int_0^{\tau} S^{a=0}(t)dt. \label{estimand-2}
\end{align}

One challenging issue for analyzing time-to-event data is right-censoring\cite{kalbfleisch2002statistical}, which may depend on covariates. Another challenging issue is confounding due to covariates imbalance in an RCT or due to non-randomization in a cohort study. To address these two challenges, a version of the targeted minimum loss estimator (TMLE) based on pseudo-observations\cite{andersen2004regression} is proposed to estimate the estimand efficiently. Furthermore, a version of the copy reference (CR) approach\cite{fang2022sequential, o2014clinical} is developed to perform a sensitivity analysis on the issue of right-censoring. 

The remainder of the paper is structured as follows. In Section 2, the proposed TMLE is described to efficiently estimate the target estimand (\ref{estimand}). In Section 3, the proposed CR approach is described for conducting a sensitivity analysis on the issue of right-censoring.  In Section 4, a real data application is provided as a tutorial on using the proposed methods. The paper is concluded with a brief summary in Section 5. 

\section{Estimator}

\subsection{Pseudo-observations for jackknife estimate}

Let $O_i, i=1, \dots, n,$ be independent and identically distributed observations. Let $\mu$ be an estimand and $\widehat{\mu}$ be the full-sample estimate based on the entire sample of size $n$. Let $\widehat{\mu}(-i)$ be the corresponding estimate based on the leave-one-out sample excluding observation $O_i$, that is, $\{O_1, \dots, O_{i-1}, O_{i+1}, \dots, O_n\}$. 

Tukey (1958) defined pseudo-observations\cite{tukey1958abstracts}, 
\begin{align}
    \widehat{\mu}_i=n\widehat{\mu}-(n-1)\widehat{\mu}(-i),\label{pv-jk}
\end{align}
for $i=1, \dots, n$. Then the jackknife estimate can be expressed as the mean of $n$ pseudo-observations, 
\begin{align}
    \widehat{\mu}_{jack}=\frac{1}{n}\sum_{i=1}^n\widehat{\mu}_i,
\end{align}
and the variance of $\widehat{\mu}_{jack}$ can be estimated by 
\begin{align}
    \widehat{Var}(\widehat{\mu}_{jack})=\frac{1}{n(n-1)}\sum_{i=1}^{n}(\widehat{\mu}_i-\widehat{\mu}_{jack})^2.
\end{align}

\subsection{Pseudo-observations for RMST estimate}

Assume the dataset consists of $n$ observations, with the $i$th observation being $O_i=(X_{i}, A_i, Y_i=\min(T_i, C_i), \delta_i)$, $i=1, \dots, n$, where $X_i$ is the vector of baseline covariates, $T_i$ is the time-to-event, $C_i$ is the censoring time, $Y_i=\min(T_i, C_i)$ is the observed outcome, and $\delta_i$ is the indicator of event occurrence (1 for the event and 0 for censoring). Arrange the data in two groups (arms or cohorts): $\mathcal{O}_{a}=\{O_{a,i}=(X_{ai}, Y_{ai}, \delta_{ai}), i=1, \dots, n_a\}$ for the group of $n_a$ subjects treated with treatment $a$, where $a=1$ or $0$ and $n=n_1+n_0$.

To define pseudo-observations, Andersen et al.~(2004) proposed to consider the Kaplan--Meier estimate, $\widehat{S}^a(t)$, for the survival function $S^a(t)$ based on the data $\mathcal{O}_a$. That is, 
\begin{align}
    \widehat{\mu}_a=\int_0^{\tau}\widehat{S}^a(t)dt.
\end{align}
Consequently, let $\widehat{\mu}_a(-i)$ be the corresponding estimate based on the data excluding $O_{a,i}$ from $\mathcal{O}_a$, that is, $\{O_{a,1}, \dots, O_{a,i-1}, O_{a,i+1}, \dots, O_{a, n_a}\}$ Thus, pseudo-observations are defined as: 
\begin{align}
    P_{ai}=n_a \widehat{\mu}_a-(n_a-1) \widehat{\mu}_a(-i),
\end{align}
where $a=1, 0$ and $i=1, \dots, n_a$. 

The pseudo-observations can be output from the SAS procedure RMSTREG performed to $\mathcal{O}_1$ and $\mathcal{O}_0$, respectively.  

\subsection{Doubly robust methods}

After the pseudo-observations are obtained for each treatment group, they can be merged as one dataset, $\mathcal{PO}=\{(X_i, A_i, P_i), i=1, \dots n\}$. Using the theory of generalized estimating equations (GEE), Andersen et al.~(2004) showed that ``once the pseudo-observations are computed, the estimation of [the treatment effects] can be carried out using standard statistical software like SAS’s PROC GENMOD." Released in 2018, RPOC RMSTREG can implement these two steps in one procedure---the calculation of the pseudo-observations in the first step and the implementation of GEE in the second step.  

However, for an RCT with imbalanced covariates and an observational cohort study, doubly robust methods \cite{hernan2020causal} are desirable to estimate the treatment effect $\theta$. Double robustness means that the estimator is asymptotically consistent if either the outcome regression model (for modeling $P_i\sim X_i + A_i$) or the propensity score model (for modeling $A_i\sim X_i$) is correctly specified. 

There are two well-known doubly robust methods in the literature on causal inference, Augmented Inverse Probability of Treatment Weighting (AIPTW)\cite{hernan2020causal} and Targeted Maximum Likelihood Estimation (TMLE)\cite{van2011targeted}. 

To implement AIPTW, a similar statement can be made to that stated in Andersen et al.~(2004) can be made: 
once the pseudo-observations are computed, the estimation of the treatment effect by the AIPTW method for the adjustment of covariates can be carried out using standard statistical software such as SAS's PROC CAUSALTRT. 

Furthermore, to improve the robustness and efficiency of AIPTW, the TMLE method can be applied. In the AIPTW method, a generalized linear model (GLM) is specified for the outcome regression model and another GLM is specified for the propensity score model, and double robustness means that if at least one of them is correctly specified, then the resulting estimator is asymptotically consistent.  In the TMLE method, a super learner model\cite{van2007super} is specified for the outcome regression model and another super learner model is specified for the propensity score model, and double robustness means that if at least one of them is asymptotically consistent, then the resulting estimator is asymptotically consistent; moreover, if both models are asymptotically consistent, then the resulting estimator is efficient. Since a Super Learner model is more likely to be consistent than a GLM, TMLE could improve the robustness and efficiency compared to AIPTW. 

To implement TMLE, a similar statement as the one stated in Andersen et al.~(2004) can be made: 
once the pseudo-observations are computed, the estimation of the treatment effect by the TMLE method for covariates adjustment can be carried out using standard statistical software like SAS's PROC CAEFFECT and R package ``tmle." \cite{gruber2012tmle}

\section{Sensitivity Analysis}

As a key step in the estimand framework \cite{ich2020e9}, sensitivity analysis is needed ``with the intent to explore the robustness of inferences from the main estimator to deviations from its underlying modeling assumptions and limitations in the data." An underlying assumption made for analyzing the time-to-event data with right-censoring is that the probability of censoring depends on the baseline covariates and treatment assignment---referred to as censoring at random (CAR) in this paper if the terminology of missing at random (MAR) is adopted. 

The literature on sensitivity analysis for missing at random (MAR) is extensive \cite{mallinckrodt2020estimands}. In this section, motivated by the idea of a copy-reference method from the literature on MAR, a version of the copy-reference method is proposed to conduct sensitivity analysis if the CAR assumption is violated.   

The copy-reference method is one of a group of reference-based methods. It invisages a hypothetical scenario of censoring not at random (CNAR) in which the censored observations in the treated arm ($A=1$) would have followed the same pattern as the one observed in the control arm ($A=0$). This can be implemented through the following four steps:

Step 1: For each treatment group, according to the original data $\mathcal{O}_{a}=\{O_{a,i}=(X_{ai}, Y_{ai}, \delta_{ai}), i=1, \dots, n_a\}$, obtain a set of pseudo-observations $\{P_{ai}, i=1,\dots,n_a\}$, $a=1, 0$. 

Step 2: Merging the subset of the observations of the treated group with $\delta_{1i}=0$ with the whole set of the observations of the control group, a new set of observations is created, that is, 
\begin{align}
    \mathcal{O}'_{0}=\{(X_{1i}, Y_{1i}, \delta_{1i}=0)\}\cup \mathcal{O}_0.
\end{align}

Step 3: According to the new dataset $\mathcal{O}'_{0}$, obtain a set of pseudo-observations, the cardinality of which is equal to the cardinality of $\mathcal{O}'_0$. 

Step 4: In the treated group, for each subject with $\delta_{1i}=0$, replace the subject's pseudo-observation $P_{1i}$ obtained in Step 1 by the corresponding pseudo-observation, denoted as $P'_{1i}$, obtained in Step 3, resulting in an updated set of pseudo-observations, $\{P'_{1i}, i=1, \dots, n_1\}$, where $P'_{1i}=P_{1i}$ if $\delta_{1i}=1$. 

Step 5: Based on the updated set of pseudo-observations for the treated arm, $\{P'_{1i}, i=1, \dots, n_1\}$, and the original set of pseudo-observations for the control arm, $\{P_{0i}, i=1, \dots, n_0\}$, obtain a new estimate, called the CR estimate, for each of the estimators discussed in Section 2. 

If the CR estimate were to be similar to the estimate of the main estimator, then it would suggest that the main result is robust to the CAR assumption, at least under the CR scenario.

\section{Tutorial}

In this section, we illustrate the proposed methods using data from the AIDS clinical trial referred to as ``ACTG175."\cite{hammer1996trial} This dataset is
available in the R package ``speff2trial" on https://cran.r-project.org.
ACTG175 was a randomized trial that evaluated antiretroviral therapy regimens among HIV-1 infected participants.\cite{hammer1996trial} Randomization was stratified according to the length of previous antiretroviral therapy (naive or experienced). 

We considered two treatment arms, zidovudine plus didanosine ($A=1$; $n_1=522$ subjects) and zidovudine ($A=0$; $n_0=532$ subjects). The outcome variable $Y$ was the time in days from randomization to the development of acquired immunodeficiency syndrome (AIDS), death, or disease progression. We considered five baseline covariates: baseline CD4 cell count ($X_1$), age ($X_2$), weight ($X_3$), sex ($X_4$), and the stratification factor ($X_5$). If the cutoff time $\tau=160$ days, the estimand of interest is the RMST at $\tau=160$.

\subsection{Main estimate using pseudo-observations and TMLE}

The main estimator contains two steps: (1) produce pseudo-observations, and (2) apply the treatment estimation method. For the first step, we can apply SAS PROC RMSTREG to produce pseudo-observations $P_{ai}, i=1, \dots, n_{ai}, a=1,0$, as shown in the last column of Table \ref{table1}.  

\FloatBarrier
\begin{table}[!htbp]
\begin{center}
  \caption{\label{table1} A partial dataset with pseudo-observations in the last column}
\begin{tabular}{ccc ccc ccc c}
    \hline
 ID & $A$ & $X_1$ & $X_2$ & $X_3$ & $X_4$ & $X_5$  & $Y$ & $\delta$ & $P$ \\
 \hline
 \hline
1 & 1 & 235 & 46 & 88.9 & M & 1 & 169 & 0 & 161.16\\
2 & 1 & 444 & 30 & 50.8 & F & 0 & 68 & 0 & 151.36\\
\dots & \\
521 & 1 & 739 & 36 & 60 & M & 0 & 95 & 1 & 90.23\\
522 & 1 & 483 & 20 & 72.4 & F & 0 & 135 & 0 & 160.32\\
\hline
523 & 0 & 504 & 43 & 66.7 & M & 1 & 156 & 0 & 162.67\\
524 & 0 & 244 & 31 & 73.0 & M & 1 & 113 & 1 & 107.97\\
\dots & \\
1053 & 0 & 373 & 17 & 103.0 & M & 1 & 56 & 0 & 142.75\\
1054 & 0 & 166 & 14 & 60    & M & 0 & 66 & 1 & 60.50\\
\hline
\multicolumn{10}{l}{{\footnotesize Note: $P$ is the pseudo-observation from PROC RMSTREG ($\tau=160$)}}
\end{tabular}
\end{center}
\end{table}

For the second step, we apply a prespecified treatment estimation method (say, TMLE using the R package ``tmle") to analyze the data $\mathcal{PO}=\{(X_i, A_i, P_i), i=1, \dots, n\}$. The TMLE method gives us the point estimate of the treatment effect, along with its standard error estimate, p-value, and confidence interval. As a result, the TMLE estimate is $\widehat{\theta}_{\mbox{\tiny TMLE}}=16.7$, with the standard error estimated as $\mbox{SE}=5.48$.

\subsection{Sensitivity analysis under the CR scenario}

To explore the robustness of the above main result under the copy-reference scenario, we recreate a tentative dataset consisting of those censored subjects in the treated arm and all subjects in the control arm. Based on the tentative dataset, we can apply SAS PROC RMSTREG to produce pseudo-observations, as shown in the last column of Table \ref{table2}. 

It is important to note that in Table \ref{table1} pseudo-observations are calculated for the two arms respectively, while in Table \ref{table2} pseudo-observations are calculated as if all the subjects in the tentative dataset were from a single control arm.

\begin{table}[ht!]
\begin{center}
  \caption{\label{table2} The tentative dataset for copy-reference imputation}
\begin{tabular}{ccc ccc ccc c}
    \hline
 ID & $A$ & $X_1$ & $X_2$ & $X_3$ & $X_4$ & $X_5$ & $Y$ & $\delta$ & Tentative $P$ \\
 \hline
 \hline
1 & 1 & 235 & 46 & 88.9 & M & 1 & 169 & 0 & 161.24\\
2 & 1 & 444 & 30 & 50.8 & F & 0 & 68 & 0 & 153.18\\
\dots & \\
522 & 1 & 483 & 20 & 72.4 & F & 0 & 135 & 0 & 160.90\\
\hline
524 & 0 & 504 & 43 & 66.7 & M & 1 & 156 & 0 & 161.24\\
525 & 0 & 244 & 31 & 73.0 & M & 1 & 113 & 1 & 107.20\\
\dots & \\
1053 & 0 & 373 & 17 & 103.0 & M & 1 & 56 & 0 & 151.00\\
1054 & 0 & 166 & 14 & 60    & M & 0 & 66 & 1 & 60.42\\
\hline
\multicolumn{10}{l}{{\footnotesize Note: Tentative dataset---censored subjects in $A=1$ plus subjects in $A=0$}}
\end{tabular}
\end{center}
\end{table}

Now we can combine the above two tables to obtain Table \ref{table3}. Simply put, in Table \ref{table1}, we keep the pseudo-observations for those subjects with $A=1$ and $\delta=1$ and for those subjects with $A=0$, and only update the pseudo-observations for those subjects with $A=1$ and $\delta=0$ using the corresponding tentative pseudo-observations in Table \ref{table2}.

\begin{table}[ht!]
\begin{center}
  \caption{\label{table3} The updated dataset after copy-reference imputation}
\begin{tabular}{ccc ccc ccc c}
    \hline\hline
 ID & $A$ & $X_1$ & $X_2$ & $X_3$ & $X_4$ & $X_5$ & $Y$ & $\delta$ & Updated $P$ \\

 \hline
1 & 1 & 235 & 46 & 88.9 & M & 1 & 169 & 0 & 161.24\\
2 & 1 & 444 & 30 & 50.8 & F & 0 & 68 & 0 & 153.18\\
\dots & \\
521 & 1 & 739 & 36 & 60 &   M & 0 & 95 & 1 & 90.23\\
522 & 1 & 483 & 20 & 72.4 & F & 0 & 135 & 0 & 160.90\\
\hline
524 & 0 & 504 & 43 & 66.7 & M & 1 & 156 & 0 & 162.27\\
525 & 0 & 244 & 31 & 73.0 & M & 1 & 113 & 1 & 107.97\\
\dots & \\
1053 & 0 & 373 & 17 & 103.0 & M & 1 & 56 & 0 & 142.75\\
1054 & 0 & 166 & 14 & 60    & M & 0 & 66 & 1 & 60.50\\
\hline
\end{tabular}
\end{center}
\end{table}

Finally, based on the updated pseudo-observations in Table \ref{table3}, we can apply the same TMLE methods to analyze the data $\mathcal{PO}=\{(X_i, A_i, P'_i), i=1, \dots, n\}$. The TMLE method gives us the updated results under the CR scenario: the TMLE estimate is $\widehat{\theta}_{\mbox{\tiny TMLE}}=16.6$, with the standard error estimated as $\mbox{SE}=5.90$. Comparing the main results and these new results under the CR scenario, we conclude that the results are robust under either the CAR assumption and the CR scenario.

\section{Summary}

Using the notion of pseudo-observations \cite{andersen2004regression}, a targeted learning framework is proposed to analyze survival data. The framework is aligned with the estimand framework of ICH E9(R1), consisting of three key steps. First, define the targeted estimand in terms of RMST. Second, construct the TMLE estimator with the pseudo-observations being the outcome variable. Third, perform some sensitivity analysis to explore the robustness of the main results if the CAR assumption is violated.   

\section*{Disclosure}
This manuscript was supported by AbbVie, and AbbVie participated in the review and approval of the content. The authors are AbbVie employees and may own AbbVie stock.


\bibliographystyle{wileyNJD-AMA}
\bibliography{references}
\end{document}